\documentclass[twocolumn,english,aps,pra,10pt,showpacs,amsmath,floatfix,superscriptaddress]{revtex4-1}
\usepackage{graphicx}
\usepackage{amsthm}
\usepackage{amssymb}
\usepackage{braket}
\usepackage[usenames,dvipsnames,svgnames,table]{xcolor}
\usepackage[unicode=true,
            pdfusetitle,
            bookmarks=true,
            bookmarksnumbered=false,
            bookmarksopen=false,
            breaklinks=true,
            pdfborder={0 0 0},
            backref=false,
            colorlinks=true]{hyperref}
\hypersetup{linkcolor=NavyBlue,urlcolor=NavyBlue,citecolor=NavyBlue}

\renewcommand\Re{\operatorname{Re}}
\renewcommand\Im{\operatorname{Im}}
\newcommand{\id}{I}
\newcommand{\bk}[1]{\left(#1\right)}
\newcommand{\Bk}[1]{\left[#1\right]}

\DeclareMathOperator{\tr}{Tr}

\begin{document}

\title{Quantum Weiss-Weinstein bounds for quantum metrology}

\author{Xiao-Ming Lu}
\email{luxiaoming@gmail.com}
\affiliation{Department of Electrical and Computer Engineering, National University of Singapore, 4 Engineering Drive 3, Singapore 117583}

\author{Mankei Tsang}
\email{mankei@nus.edu.sg}
\affiliation{Department of Electrical and Computer Engineering, National University of Singapore, 4 Engineering Drive 3, Singapore 117583}
\affiliation{Department of Physics, National University of Singapore, 2 Science Drive 3, Singapore 117551}

\begin{abstract}
  Sensing and imaging are among the most important applications of
  quantum information science. To investigate their fundamental limits
  and the possibility of quantum enhancements, researchers have for
  decades relied on the quantum Cram\'er-Rao lower error bounds
  pioneered by Helstrom. Recent work, however, has called into
  question the tightness of those bounds for highly nonclassical
  states in the non-asymptotic regime, and better methods are now
  needed to assess the attainable quantum limits in reality. Here we
  propose a new class of quantum bounds called quantum Weiss-Weinstein
  bounds, which include Cram\'er-Rao-type inequalities as special
  cases but can also be significantly tighter to the attainable
  error. We demonstrate the superiority of our bounds through the
  derivation of a Heisenberg limit and phase-estimation examples.
\end{abstract}

\maketitle

\section{Introduction}
Quantum noise is becoming a major limiting factor in sensing and
imaging technology, with photon shot noise in particular imposing
limits to modern gravitational-wave detectors \cite{ligo16c} as well
as optical microscopes \cite{pawley}. Quantum metrology
\cite{Giovannetti2004,Giovannetti2011}, through the use of
nonclassical states or innovative measurement schemes, promises to
beat such conventional quantum limits and offer significant accuracy
enhancements. This promise has led to renewed interest in the quantum
estimation theory pioneered by Helstrom \cite{Helstrom1976}, and
especially the quantum Cram\'er-Rao bounds (QCRBs)
\cite{Helstrom1976,Holevo1982,Yuen1973}. The bounds were originally
developed to investigate thermal and laser sources, but they are now
being applied to increasingly exotic quantum states for the purpose of
quantum enhancements. The asymptotic attainability of Helstrom's QCRB
for one parameter \cite{hayashi05,fujiwara2006} and Holevo's version
for multiple parameters \cite{gill_guta} suggests that the QCRBs can
be tight; many proposals of quantum enhancements are based just on
QCRBs~\cite{Anisimov2010,Rivas2012a,Zhang2013a,humphreys}. Unfortunately,
these works ignore the number of repeated trials needed to reach the
asymptotic regime, and the requirement of many repetitions can negate
the perceived advantage of their protocols, as realized by subsequent
studies
\cite{Tsang2012,Berry2015,Zhang2014,Nair,Giovannetti2012,Giovannetti2012a,Hall2012,Hall2012a,Hall2012b}.

These mishaps suggest that Helstrom's paradigm of quantum estimation
theory can no longer fulfill the modern demands of quantum metrology
and better approaches are needed to assess quantum sensors in highly
nonclassical states.  We can take inspiration from classical
estimation theory, where it is common knowledge that Cram\'er-Rao-type
inequalities can grossly underestimate the attainable estimation error
\cite{VanTrees2007}.  Two new families of bounds have emerged there as
the best candidates to supersede the Cram\'er-Rao family
\cite{VanTrees2007}: the Ziv-Zakai family \cite{ziv1969} and the
Weiss-Weinstein family \cite{Weiss1985,Weinstein1988}.  Although they
are derived from distinct principles, both have been found to remain
remarkably tight to the attainable estimation errors in both
non-asymptotic and asymptotic regimes, with diverse applications in
engineering \cite{VanTrees2007} as well as astronomy
\cite{Nicholson1998}. For the quantum problem, quantum Ziv-Zakai
bounds (QZZBs) have recently been proposed and shown to be superior to
QCRBs in many cases \cite{Tsang2012,Berry2015,Zhang2014}.  Although
the QZZBs are trivial to prove and straightforward to evaluate, there
is no general guarantee about their superiority over the QCRBs, so
they have to be compared on a case-by-case basis. To overcome this
problem, here we propose quantum versions of the Weiss-Weinstein
bounds, which have the advantage of including Cram\'er-Rao-type
inequalities as special cases. Through the derivation of a Heisenberg
limit and examples of phase estimation, we further demonstrate that
our new bounds can not only beat QCRBs but also QZZBs for tightness.

\section{Results}

\subsection{Quantum covariance inequality}
Our quantum Weiss-Weinstein bounds (QWWBs) are based on a quantum
generalization of the covariance inequality proposed by Weinstein and
Weiss \cite{Weinstein1988,VanTrees2007}. It is a lower bound on the 
global estimation error matrix defined as
\begin{align}
  \Sigma:=\int\!dx dy\,\epsilon(x,y)\epsilon(x,y)^{\top}p(x,y),
\end{align}
where $x \in \mathbb R^J$ is a column vector of $J$ unknown
parameters, $y$ is the observation, $p(x,y)$ is their joint
probability distribution, $\epsilon(x,y) := \tilde x(y)-x$ is the
error vector with respect to an estimator $\tilde x(y)$, and $\top$
denotes the matrix transpose. For the quantum problem \cite{Yuen1973},
\begin{align}
  p(x,y) &= \tr[E_y \rho(x)],
\end{align}
where $\rho(x) = \rho_x p(x)$ is the hybrid density operator
\cite{Tsang2011}, $\rho_x$ is the conditional density operator that
models the quantum system as a function of $x$, $p(x)$ is the prior
distribution, $E_y$ is the positive operator-valued measure (POVM)
that models the quantum measurement \cite{Helstrom1976,Holevo1982},
and $\tr$ denotes the operator trace.  Our quantum covariance
inequality reads
\begin{align} \label{eq:covariance_inequality}
    \mathrm{\Sigma}\geq CG^{-1}C^{\top},
\end{align}
where $G$ is a $K\times K$ real and strictly positive matrix defined
as
\begin{align} \label{eq:def_G}
    G_{kk'} := \int\!dx\,\Re\tr\Bk{L_k(x)^\dagger L_{k'}(x) \rho(x)}
\end{align}
in terms of a set of operators $\{L_k(x); k = 1,2,\dots,K\}$ and $C$
is a $J\times K$ real matrix defined as
\begin{align} \label{eq:def_C}
    C_{jk} := \int\!dx dy\,\epsilon_j(x,y) \Re\tr\Bk{E_y L_k(x )\rho(x)}.
\end{align}
Equation~(\ref{eq:covariance_inequality}) means that
$\Sigma - CG^{-1}C^{\top}$ is positive-semidefinite.  The proof of
Eq.~(\ref{eq:covariance_inequality}) is given in the Methods.  To
derive measurement-independent quantum bounds, we will choose a set of
$L_k(x)$'s to make $C$ independent of the POVM and the estimator.

\subsection{Quantum Weiss-Weinstein bounds}
Our QWWBs posit that each $L_k(x)$ satisfies
\begin{align} \label{eq:L_k}
D_k(x)
&=   \frac{1}{2}\Bk{L_k (x) \rho(x) + \rho(x) L_k(x)^{\dagger}},
\\
\label{eq:D_k}
D_k(x) &:= \frac{V_k(x+h_k)-V_k(x)}{|h_k|},
\\
\label{eq:V_k}
V_k(x)&:=  \mathcal{N}_k\,\rho(x )^{s_k } \circ \rho(x -h_k )^{1-s_k },
\end{align}
where $h_k$ is a real vector with length $|h_k|$ and the same
dimension as that of $x$, $0<s_k <1$,
$O_1\circ O_2 := (O_1 O_2 + O_2 O_1)/2$ denotes the Jordan product,
and $\mathcal{N}_k$ is a normalization factor such that
$\int\!dx\,\tr V_k(x) = 1$.  This choice of $L_k(x)$ gives
\begin{align}
C_{jk}=\frac{h_{kj}}{|h_k|},
\label{C}
\end{align}
where $h_{kj}$ is the $j$th component of $h_k$.  To see this, notice
that, with $s_k$ being set in the range $(0,1)$, $V_k(x)$ vanishes
where $p(x)$ vanishes, leading to
$\int\!dx\, V_k(x+h) = \int\!dx\, V_k(x)$ and $\int\!dx\, D_k(x) = 0$,
as the domain of integration is $\mathbb R^J$ and $p(x)$ must vanish
at infinity. It then follows from Eqs.~(\ref{eq:def_C}) and
(\ref{eq:L_k}) and the completeness property of $E_y$ that
$C_{jk} = -\int\!dx\,x_j\tr [V_k(x+h)-V_k(x)]/|h_k|$.  A change of
variables gives
$\int\!dx\,x_j V_k(x+h_k) = \int\!dx\,(x_j-h_{kj}) V_k(x)$, which
leads to Eq.~(\ref{C}).


The QWWBs given by Eqs.~(\ref{eq:covariance_inequality})--(\ref{C})
are applicable to any quantum measurement, any biased or unbiased
estimator, and do not require $\rho_x$ or $p(x)$ to be
differentiable. They are a family of bounds that hold for any $K$, any
$h_k$, and any $0<s_k<1$, such that tighter versions can be obtained
by choosing these parameters judiciously. The $|h_k|\to 0$ limit
leads to the Bayesian QCRBs \cite{Yuen1973,Tsang2011} (see Appendix~\ref{app_a}), while finite $h_k$ and $s_k \to 1$
lead to Bayesian multiparameter versions of the quantum bounds
proposed by Tsuda and Matsumoto \cite{Tsuda2005}.  
The classical Weiss-Weinstein bound is usually computed with $s_k=1/2$ since it often maximizes the bound~\cite{Weiss1985,Weinstein1988,VanTrees2007}; our examples later show that $s_k = 1/2$ can also lead to tight quantum bounds.

The $L_k(x)$ operators may not be uniquely determined by
Eq.~(\ref{eq:L_k}) for a given $\rho(x)$ and $D_k(x)$.  We prove in
Appendix~\ref{app_b} that the Hermitian
$L_k(x)$'s give the tightest QWWB, though non-Hermitian choices may be
easier to obtain in some cases. 
When $\rho(x)$ and $D_k(x)$ are of low rank, the following expression is useful to obtain the Hermitian $L_k(x)$'s:
\begin{equation}\label{eq:solution}
	L_k(x) = \sum_{\alpha,\beta\mid\lambda_\alpha+\lambda_\beta\neq0} 
	\frac{2\braket{\alpha|D_k(x)|\beta}}{\lambda_\alpha+\lambda_\beta} \ket\alpha\bra\beta,
\end{equation}
where each $\ket\alpha$ is an eigenstate of $\rho(x)$ with eigenvalue $\lambda_\alpha$.
Taking $D_k(x)$ as the partial derivative with respect to $x_k$, Eq.~(\ref{eq:solution}) is a well-known expression for the symmetric logarithmic derivative operator~\cite{Helstrom1976,Braunstein1994}.
For non-Hermitian $L_k(x)$'s, the QWWB
can be tightened by noting that $L_k(x)+i\alpha_k$, with $\alpha_k$
being an arbitrary real number, is also a solution of
Eq.~(\ref{eq:L_k}).  Maximizing the positive matrix $G$ over
$\alpha_k$ leads to $\alpha_k = -\Im\langle L_k(x) \rangle$, where
$\langle\bullet\rangle:=\int\!dx\tr[\bullet\rho(x)]$.  We can
therefore always replace $G$ by $G-\Delta$ to tighten the QWWBs, where
$\Delta_{kk'}=\Im\langle L_k(x)\rangle \Im\langle L_{k'}(x)\rangle$ is
a positive-semidefinite matrix.

The QWWBs degenerates into the classical Weiss-Weinstein bounds~\cite{Weiss1985,Weinstein1988,VanTrees2007} for a commuting family of $\rho_x$.
In such a situation, we can identify a basis $\{\ket y\}$ in which all $\rho_x$ are diagonal matrices, meaning that $\rho(x)$ can be equivalently expressed as a joint probability $p(x,y):=\braket{y|\rho(x)|y}$. 
Consequently, $L_k(x)$ is also diagonal with the basis $\{\ket y\}$, and can be expressed as a function
\begin{equation}
	L_k(x,y) = \frac{\mathcal{N}_k}{|h_k|}\Big\{
	\Big[\frac{p(x+h_k,y)}{p(x,y)}\Big]^{s_k} -
	\Big[\frac{p(x-h_k,y)}{p(x,y)}\Big]^{1-s_k}\Big\},
\end{equation}
where $\mathcal{N}_k= \mathbb{E}[p(x+h_k,y)^{s_k}/p(x,y)^{s_k}]^{-1}$, and $\mathbb{E}[\bullet]$ denotes the expectation value with respect to the joint probability $p(x,y)$.
The Classical Weiss-Weinstein bound is still of the form Eq.~(\ref{eq:covariance_inequality}) with $C$ being given by Eq.~(\ref{C}), whereas $G$ is expressed in a classical manner as  
$G_{kk'} = \mathbb{E}[L_k(x,y)L_{k'}(x,y)]$.

\subsection{Single-parameter estimation}
For single-parameter estimation, the error matrix reduces to the
mean-square error $\Sigma=\int\!dx dy\,[\tilde x(y)-x]^2 p(x,y)$.  The
QWWBs become
\begin{align}\label{eq:general_QWWB}
    \Sigma\geq\Sigma_\mathrm{W}(s,h):=\frac{1}{\langle L(x)^{\dagger}L(x)\rangle-[\Im\langle L(x)\rangle]^{2}}.
\end{align}
The following choice of $L(x)$ serves our purpose:
\begin{align}\label{eq:special_L}
    L(x)=\frac{\mathcal{N}(s,h)}{|h|}\left[\Lambda(s,h) - \Lambda(1-s,-h)\right],
\end{align}
where $\Lambda(s,h):=\rho(x+h)^s\rho(x)^{-s}$ and
$\mathcal{N}(s,h)=\langle\Lambda(s,h)\rangle^{-1}$.  Here we use the
convention that a power of a positive-semidefinite operator is taken
only on its support~\cite{Hiai2011}.  $\rho(x)^{-1}$ is then the
generalized inverse defined on the support and $\rho(x)^{0}$ is the
projector onto the support. Consequently, $L(x)$ vanishes where $p(x)$
vanishes for $0<s<1$.  Equation~(\ref{eq:general_QWWB}) becomes
\begin{align}\label{eq:QWWB}
    \Sigma_{\mathrm{W}}(s,h) = \frac{h^2 g(s,h)^2}{ g(2s,h) + g(2-2s,-h) - 2\tilde{g}(s,2h)},
\end{align}
where $g(s,h) := \langle\Lambda(s,h)\rangle$ and $\tilde g(s,2h) := \Re\braket{\Lambda(s,h)^\dagger\Lambda(1-s,-h)}$.
When the conditional density operators $\rho_x$ are of full rank, it can be shown that $\tilde{g}(s,h)=g(s,h)$.
Equation~(\ref{eq:QWWB}) is then of the same form as the classical Weiss-Weinstein bound~\cite{Weiss1985}, but with a different function $g(s,h)$.
Although the characteristics of $g(s,h)$ determines the QWWB in an intricate manner, some intuitive observations can be given as follows.
The situation of particular interest is that $\Sigma_{\rm W}(s,h)$ takes its maximum at a finite large value of $h$ rather than at $h\to0$, meaning that the QCRB underestimates the error.
For the case of $\tilde g(s,h)=g(s,h)$, the denominator of Eq.~(\ref{eq:QWWB}) is bounded above by $2$ due to $g(s,h)\in[0,1]$.
Then, considering the factor $h^2$ in the numerator, Eq.~(\ref{eq:QWWB}) may take its maximum at a finite large value of $h$, when $g(s,h)$ is not always far less than one as $h$ becomes large.  
The estimation models with such a characteristic of $g(s,h)$ may be poorly assessed by only the QCRB, thereby are in need of the QWWB or the QZZB.

We now focus on phase estimation, a paradigmatic problem in quantum
metrology.  Assume $\rho_x=\exp(-ixH)\rho\exp(ixH)$, where $\rho$ is
the initial state and $H$ is an Hermitian operator. In this case,
$g(s,h)$ and $\tilde g(s,h)$ can be neatly separated as
$g(s,h)=g_{\mathrm{c}}(s,h) g_{\mathrm{q}}(s,h)$ and
$\tilde{g}(s,h)=g_{\mathrm{c}}(s,h) \tilde{g}_{\mathrm{q}}(s,h)$,
where
\begin{align}
g_{\mathrm{c}}(s,h) &= \int_{\{x;p(x)>0\}}\!\!dx\,p(x+h)^s p(x)^{1-s}
\end{align}
is a classical component that depends only on the prior, and
\begin{align}
  g_{\mathrm{q}}(s,h) &= \tr(\rho_h^s \rho^{1-s}),\\
  \tilde{g}_{\mathrm{q}}(s,2h) &= \Re\tr(\rho_h^s \rho_{-h}^{1-s}\rho^0)
\end{align}
are quantum components. If the initial state is pure,
$\rho=\ket{\psi}\langle\psi|$, and since $\rho^s=\rho$ for a pure
state, we obtain $g_\mathrm{q}(s,h)=|z(h)|^2$ and
$\tilde{g}_\mathrm{q}(s,2h)=\Re\,z(h)^2z(2h)^*$, where
$z(h):=\langle\psi|\exp(-ihH)\ket{\psi}$.  Interestingly, the quantity
$g_\mathrm{q}(s,h)$ also plays an important role in the quantum
Chernoff bound for binary hypothesis testing
\cite{Audenaert2007,Nussbaum2009,Audenaert2008},
although no meaningful relationship between the Weiss-Weinstein bound  and the Chernoff bound, apart from the coincidental mathematical similarity, has been discovered to our knowledge.

\subsection{Heisenberg limit}
The QWWBs can be used to derive a Heisenberg limit as follows.  Let
$\ket{\psi} = \sum_j c_j \ket{j}$ be a purification of the initial
quantum state, where each $\ket{j}$ is an eigenvector of $H$ with
eigenvalue $E_j$. Then
$\tilde g_\mathrm{q}(s,2h)=\sum_{jkl}|c_j c_k
c_l|^2\cos[h(E_j+E_k-2E_l)]$.  The cosine function can be bounded as
$\cos\theta \geq 1-\lambda|\theta|$, where $\lambda \approx 0.7246$ is
the implicit solution of $\lambda = \sin\phi = (1-\cos\phi)/\phi$
\cite{Tsang2012}.  Thus,
$\tilde g_\mathrm{q}(s,2h) \geq 1- \lambda
|h|\sum_{jkl}|c_jc_kc_l|^2|E_j+E_k-2E_l|$.  Let $E_0$ be the minimum
eigenvalue of $H$ and $\Delta E_j := E_j - E_0$.  By noting that
$|E_j+E_k-2E_l|=|\Delta E_j + \Delta E_k - 2 \Delta E_l| \leq \Delta
E_j + \Delta E_k + 2 \Delta E_l$, it follows that
$\tilde g_\mathrm{q}(s,2h)\geq 1 - 4 \lambda |h| H_+$ with
$H_+:=\tr(H\rho)-E_0$.  Consequently, $\tilde g_\mathrm{q}(s,2h)$ is
nonnegative when $|h|\leq1/(4\lambda H_+)$; this implies that the QWWB
is further bounded as
\begin{multline}\label{eq:before_H_limit}
    \Sigma\geq \Sigma_\mathrm{W}^\prime(h) := \kappa(h) h^2 |z(h)|^2 
    \mbox{ with } \\
    \kappa(h):=\sup_{0<s<1}\frac{g_\mathrm{c}(s,h)^2}{g_\mathrm{c}(2s,h)+g_\mathrm{c}(2-2s,-h)}
\end{multline}
for $|h|\leq h_\star :=1/(4\lambda H_+)$.
The quantity $|z(h)|^2$ is the quantum fidelity between $\ket{\psi}$ and $\exp(-ihH)\ket{\psi}$, which is bounded as $|z(h)|^2 \geq 1 - |2h\lambda H_+|$~\cite{Tsang2012}.
Taking $h=h_\star$, one obtains
\begin{align}\label{eq:H_limit}
    \Sigma\geq\Sigma_\mathrm{W}(h_\star)^\prime \geq \frac{\kappa(h_\star)}{32\lambda^2 H_+^2}.
\end{align}
We have not yet made any assumption about the prior, which is
incorporated in $\kappa(h_\star)$.  Since
$g_\mathrm{c}(1/2,\pm h)\leq 1$, it follows that
$\kappa(h)\geq g_\mathrm{c}(1/2,h)^2/2$.  The quantity
$g_\mathrm{c}(1/2,h)$, also known as the Bhattacharyya coefficient,
measures the overlap between the prior probability distribution $p(x)$
and its displaced version $p(x+h)$.  For a large enough $H_+$
(corresponding to a small enough $h_\star$) such that
$g_\mathrm{c}(1/2,h_\star)\approx1$, Eq.~(\ref{eq:H_limit}) gives a
Heisenberg limit as $1/(64\lambda^2H_+^2)$, which is higher than the
limit $1/(80\lambda^2H_+^2)$ derived from a QZZB in
Ref.~\cite{Tsang2012}.  Both this work and Ref.~\cite{Tsang2012} use a
linear lower bound on the fidelity; an even tighter Heisenberg limit
can be obtained via the stronger fidelity bound in
Ref.~\cite{Giovannetti2012a}.
For a generator $H$ with integer eigenvalues, a stronger Heisenberg limit was derived through some information-theoretic inequalities~\cite{Hall2012,Hall2012a}.

\subsection{Phase-estimation examples}\label{sec:two_level}
\begin{figure}[tbp!] 
    \centering
    \includegraphics[width=8cm]{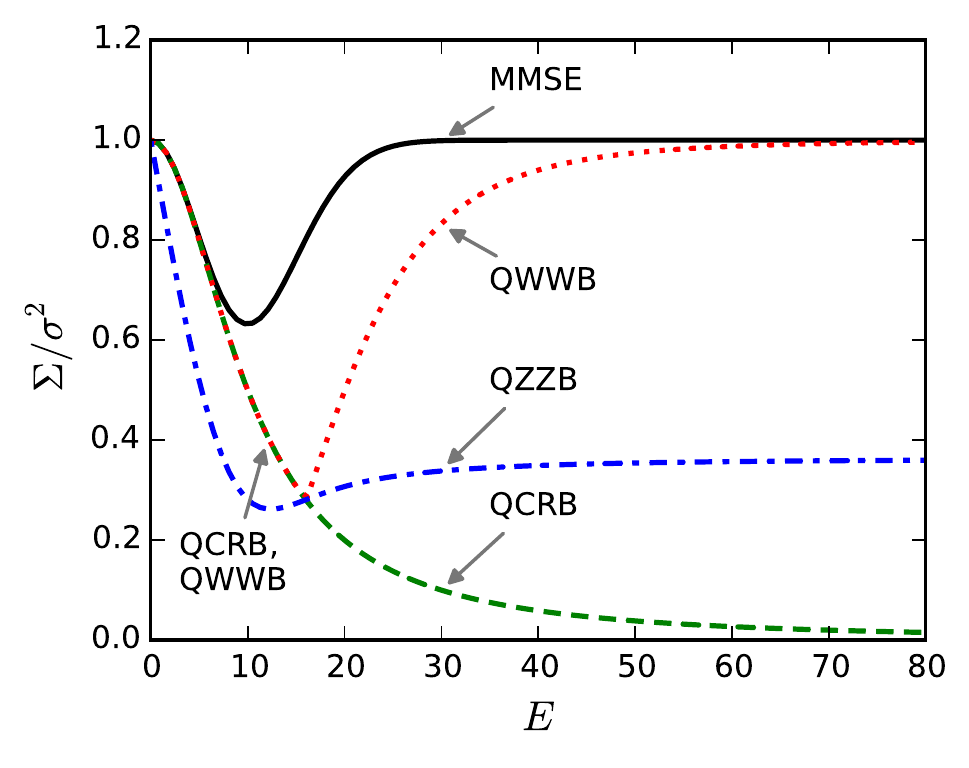}
    \caption{\label{fig:1} Comparison of the MMSE (black solid), QWWB
      (red dotted), QZZB (blue dash-dotted), and the Bayesian QCRB
      (green dashed) for the estimation of a random phase via a
      qubit. The prior distribution is Gaussian with $\sigma=0.1$
      standard deviation.  The QWWB is numerically optimized over
      $h\in[0,10\sigma]$ while $s$ is set to $1/2$.  The MMSE and the
      error bounds are normalized with respect to the prior value
      $\sigma^2$.}
\end{figure}

We now demonstrate the tightness of QWWBs relative to other existing
quantum bounds through two examples.  The first example is the
estimation of a random phase with Gaussian prior via a qubit.  Assume
that the initial qubit state is
$\ket{\psi}=(|0\rangle+|1\rangle)/\sqrt2$, the generator is
$H=E|1\rangle\langle1|$ with $E>0$, and the standard deviation of the
prior is $\sigma$.  For this simple model, the minimum mean-square
error (MMSE) can be analytically
calculated~\cite{Macieszczak2014,Jarzyna2015}, and we can use it as a
benchmark for the quantum bounds.  Setting $s=1/2$, the QWWB is given
by
\begin{align}
     \Sigma \geq \sup_h \frac{h^2\exp[-h^2/(4\sigma^2)] 
\cos(hE/2)^2}{2-2\exp[-h^2/(2\sigma^2)\cos(hE)]},
\end{align}
see Appendix~\ref{app_c} for details. Since
$\exp(-ix H)\ket{\psi}$ has a period of $2\pi/E$, $x$ and $x+2\pi/E$
are fundamentally indistinguishable from any quantum measurement. This
ambiguity means that even the optimal measurement can produce an
estimate in the wrong period, leading to substantial errors. The MMSE
stays close to the prior value $\sigma^2$ as a result, as shown in
Fig.~\ref{fig:1}. The QCRB, on the other hand, is incapable of
accounting for the phase ambiguity because of its differential nature
and severely underestimates the attainable error for large $E$. The
QZZB is not much better, and the QWWB, being close to the QCRB where
it is reasonably tight and also following the MMSE for larger $E$, is
the clear winner in this benchmark example.

\begin{figure}[tbp!] 
    \centering
    \includegraphics[width=8cm]{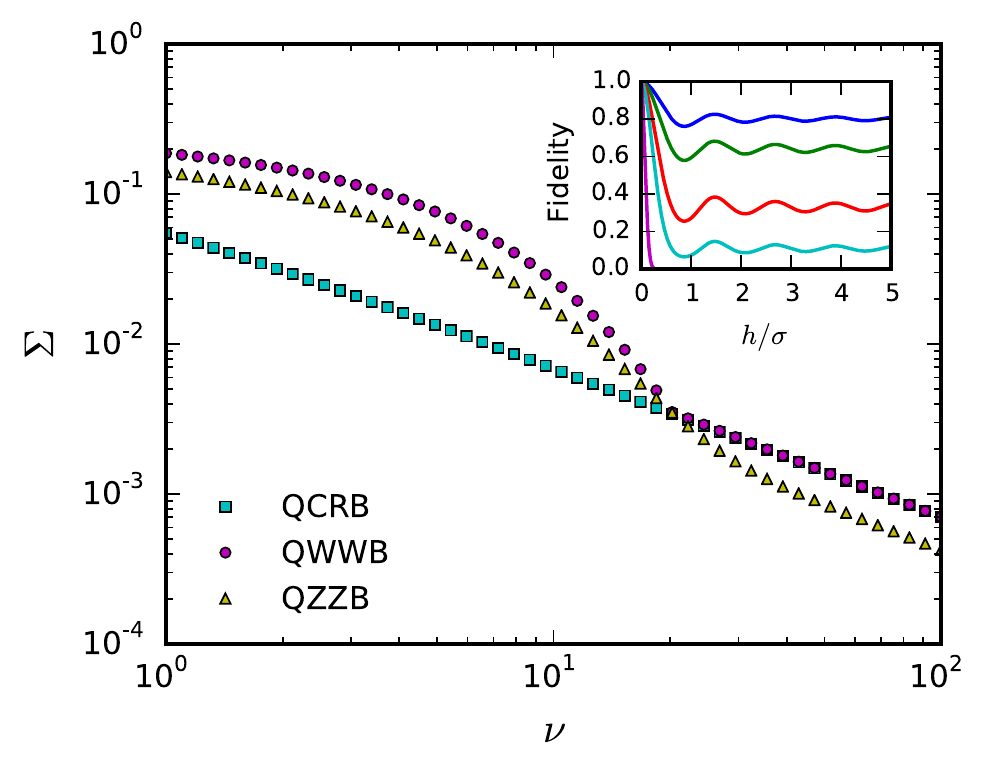}
    \caption{\label{fig:2} Error bounds versus the number $\nu$ of
      identically distributed quantum bosonic probes.  The prior
      distribution is Gaussian with $\sigma = 0.5$ standard deviation.
      The QWWB is numerically optimized over $h\in[0,10\sigma]$ with
      $s = 1/2$, and the QZZB is computed according to
      Ref.~\cite{Tsang2012}. The inset plots the fidelity
      $|\langle\psi|\exp(-ihH)\ket{\psi}|^{2\nu}$ for
      $\nu=1, 2, 5, 10, 100$ (from above to below).  }
\end{figure}

For the second example, we consider phase estimation using $\nu$
independent and identically distributed bosonic probes.  For each
probe, we assume $H=\sum_{j=0}^{\infty}j\ket{j}\bra{j}$ and
$\ket{\psi} =
\sqrt{1-\epsilon}|0\rangle+\sqrt{\epsilon/M}\sum_{j=1}^M\ket{j}$, with
$M\geq1$ being an integer and $0<\epsilon<1$ \cite{Rivas2012a}.  In
this case the MMSE is not known, and we have to rely on quantum bounds
to investigate the fundamental limit.  Figure~\ref{fig:2} compares the
three quantum bounds for $\epsilon=0.1$ and $M=10$. Though the
asymptotic attainability of the QCRB \cite{hayashi05,fujiwara2006}
means that it should be tight for large enough $\nu$, the QCRB by
itself is incapable of determining the $\nu$ needed for tightness.  It
is remarkable that the QWWB and the QZZB, though derived from
different principles, follow similar behaviors here. Both are
substantially higher than the QCRB for small $\nu$ and demonstrate a
threshold behavior as $\nu$ is increased, revealing the regime where
the prior information dominates and the QCRB is overly
optimistic. Once again, the QWWB is higher than the other bounds for
all values of $\nu$.

\subsection{Multiple test points}
Similar to the classical Weiss-Weinstein family of error
bounds~\cite{Weiss1985,Weinstein1988}, the quantum bounds can be
tightened by involving multiple test points.  As an example, 
consider the QWWB for a single-parameter estimation with two test
points, for which $G$ is a $2\times2$ matrix whose inverse can be
explicitly expressed as
\begin{equation}
	G^{-1}=\frac{1}{\det G}
	\begin{pmatrix}
		G_{22}  & -G_{12} \\
		-G_{21} &  G_{11}
	\end{pmatrix}.
\end{equation}
Since $C=(1,1)$, the lower bound from Eq.~(\ref{eq:covariance_inequality}) becomes
\begin{align}
	CG^{-1}C^\top &= \frac{G_{11} + G_{22} - G_{12} - G_{21}}{\det G}\\
	\geq & \max\left\{\frac1{G_{11}}, \frac1{G_{22}}\right\},
	\label{eq:multi_tests}
\end{align}	
where the inequality is due to the fact that $G$ is symmetric and positive, meaning that the two-test-point lower bound is tighter than that given by either of the two test points.
Following the same strategy that derives the combined Bayesian bound in classical parameter estimation~\cite{VanTrees2007}, we can set the first test point $h_1$ to an infinitesimal value and the second test point $h_2$ to a finite value, leading to a combined quantum error bound tighter than the QCRB.

\section{Discussion}
Our QWWBs set a higher standard in quantum metrology. Not only do they
include QCRBs as special cases and inherit their asymptotic tightness
at least for one parameter \cite{hayashi05,fujiwara2006}, they can
also beat the recently invented QZZBs \cite{Tsang2012} and serve as
the more natural successors of the Cram\'er-Rao family in the
post-Helstrom era of quantum metrology. Our results demonstrate that
differential geometry of quantum states alone
\cite{Braunstein1994,Bengtsson2006} cannot guarantee their usefulness;
more general distance measures, such as the quantum Chernoff distance
used in our QWWBs and the trace distance in the QZZBs, should be
consulted to establish tighter quantum limits to parameter estimation,
especially for nonclassical states or nontrivial parameter
dependence. Future proposals of quantum metrological schemes should no
longer rely only on QCRBs to support their cases without also
investigating their tightness.  We envision our QWWBs to be the new
standard against which these proposals should be assessed.

\acknowledgments
We acknowledge helpful discussions with Ranjith Nair, Shan Zheng Ang,
and Shilin Ng. This work is supported by the Singapore National
Research Foundation under NRF Grant No.\ NRF-NRFF2011-07 and Singapore
Ministry of Education Academic Research Fund Tier 1 Project
R-263-000-C06-112.

\section*{Contributions}
X.-M.~L.\ invented the quantum Weiss-Weinstein bounds presented here
and performed all the proofs and calculations.  M.~T.\ conceived the
problem.  Both authors discussed extensively during the course of this
work and contributed to the writing of the manuscript.

\appendix

\section{Proof of the quantum covariance inequality}
Here we prove Eq.~(\ref{eq:covariance_inequality}).  Let $u$
and $v$ be arbitrary real column vectors of dimension $J$ and $K$
respectively.  It follows from the definitions that
\begin{align}
    u^{\top}\Sigma u   &=  \int\!dx dy\,\epsilon_{u}(x,y)^2\tr(A^\dagger A),\\
    v^{\top}Gv         &=  \int\!dx dy\,\tr(B^\dagger B),\label{eq:vGv}
\end{align}
where $\epsilon_{u}(x,y) := \sum_j u_j \epsilon_j(x,y)$,
$A := \sqrt{E_y}\sqrt{\rho(x)}$, and
$B := \sqrt{E_y} [\sum_k v_k L_k(x)] \sqrt{\rho(x)}$.  In
Eq.~(\ref{eq:vGv}), we have used $\int\!dy\,E_y =\id$
with $\id$ being the identity operator.  As a result of the
Cauchy-Schwarz inequality,
\begin{align}
    \sqrt{u^{\top}\Sigma uv^{\top}Gv}
    \geq
    \int\!dx dy\,|\epsilon_{u}(x,y)| 
    \sqrt{\tr(A^\dagger A)\tr(B^\dagger B)}.
\end{align}
From the inequality $\tr(A^\dagger A) \tr(B^\dagger B) \geq |\tr(A^\dagger B)|^2$ followed by $|\tr(A^\dagger B)| \geq |\Re\tr(A^\dagger B)|$, we get
\begin{align}
    \bk{u^{\top}\Sigma u}\bk{v^{\top}Gv}
     \geq \left[\int\!dx dy\,\left|\epsilon_{u}(x,y) \Re\tr(A^\dagger B)\right|\right]^2 \nonumber \\
    \geq  \left|\int\!dx dy\,      \epsilon_{u}(x,y) \Re\tr(A^\dagger B)\right|^2 
     =  \left(u^{\top}Cv\right)^2. 
\end{align}
Taking $v=G^{-1}C^{\top}u$ implies
$(u^{\top}\Sigma u) (u^{\top}CG^{-1}C^{\top}u) \geq
(u^{\top}CG^{-1}C^{\top}u)^{2}$. Since $G$ is strictly positive,
$u^{\top}CG^{-1}C^{\top}u$ is positive, leading to
$u^{\top}\Sigma u\geq u^{\top}CG^{-1}C^{\top}u$.  As this inequality
holds for any real vector $u$, Eq.~(\ref{eq:covariance_inequality})
results.

\section{Relation between QWWB and QCRB} \label{app_a}

We here show that the QWWBs include the QCRB as a special case.  
Let $h_k$ be along the direction of $x_k$ in the parameter vector space.
Suppose that $\rho(x)$ is differentiable.  
When $|h_k|\to0$, one has
\begin{equation}\label{eq:approximation}
	\rho(x+h_k)^{s_k} \simeq \rho(x)^{s_k} + |h_k| \partial\rho(x)^{s_k}/\partial x_k.
\end{equation}
It can be shown from Eqs.~(\ref{eq:D_k}), (\ref{eq:V_k}) and (\ref{eq:approximation}) that 
$D_k(x) \simeq \mathcal{N}_k \partial\rho(x)/\partial x_k$, 
where the normalizing factor $\mathcal{N}_k^{-1}$ can be given by
\begin{equation}
    \mathcal{N}_k^{-1} \simeq 1 - |h_k| \int\!dx\tr \left[\rho(x)^{s_k} \frac{\partial\rho(x)^{1-s_k}}{\partial x_k}\right].
\end{equation} 
Let $\rho(x)=\sum_\alpha \lambda_\alpha |\phi_\alpha\rangle\langle\phi_\alpha|$ be the eigenvalue decomposition.
Since $\rho(x)^{s_k}=\sum_\alpha \lambda_\alpha^{s_k}|\phi_\alpha\rangle\langle\phi_\alpha|$, it follows that
\begin{align}
    &\quad \int\!dx \tr\left[\rho(x)^{s_k} \frac{\partial\rho(x)^{1-s_k}}{\partial x_k}\right] \nonumber\\
    &= \sum_\alpha \int\!dx\left[
    	\lambda_\alpha^{s_k}\frac{\partial\lambda_\alpha^{1-s_k}}{\partial x_k}
    	+ \lambda_\alpha \left(\langle\phi_\alpha|\frac{\partial\phi_\alpha}{\partial x_k}\rangle 
    	+ \langle\frac{\partial\phi_\alpha}{\partial x_k}|\phi_\alpha\rangle\right)
    \right]\nonumber\\
    &= (1-s_k)\sum_\alpha\int\!dx\frac{\partial\lambda_\alpha}{\partial x_k}
    + \sum_\alpha\int\!dx\,\lambda_\alpha \frac{\partial}{\partial x_k}\langle\phi_\alpha|\phi_\alpha\rangle\nonumber\\
    &=0,
\end{align}
where we have used $\lambda_\alpha|_{x_k=\pm\infty}=0$ in the last equality. 
Thus, $D_k(x)\simeq\partial\rho(x)/\partial x_k$ when $|h_k|\to0$. 
Consequently, the operator $L_k(x)$ becomes the symmetric logarithmic derivative operator (not necessarily to be Hermitian, see Ref.~\cite{Tsang2011}) for $\rho(x)$ with respect to $x_k$, and the resulting QWWB becomes a corresponding QCRB.

\section{Hermitian $L_k(x)$ tightening the QWWB} \label{app_b}

Here, we prove that the Hermitian $L_k(x)$ gives the tightest lower bound on the estimation-error covariance matrix among all choices of $L_k(x)$ satisfying Eq.~(\ref{eq:L_k}) for given $\rho(x)$ and $D_k(x)$.
This can be seen from the following Proposition.

\textit{Proposition.}
Suppose that $L$ is an operator satisfying 
\begin{equation}\label{supp:L}
    \frac12(L\rho+\rho L^\dagger) = D,
\end{equation}
where $\rho$ is a given positive semidefinite operator and $D$ is a given Hermitian operator.
Then, 
\begin{equation}
    \min\tr(L^\dagger L \rho ) = \tr(\tilde L^\dagger \tilde L \rho),
\end{equation} 
where the minimum is taken over all solutions of Eq.~(\ref{supp:L}) for $L$, and $\tilde L$ denotes a Hermitian solution.

\textit{Proof.}
Let $M = (L+L^\dagger)/2$ and $N=(L-L^\dagger)/(2i)$, which are both Hermitian operators. Let $\rho=\sum_j \lambda_j |j\rangle\langle j|$ be the eigenvalue decomposition.
It follows from Eq.~(\ref{supp:L}) and $L=M+iN$ that
\begin{equation}
    D_{jk} = \frac12(\lambda_k+\lambda_j)M_{jk}+\frac{i}{2}(\lambda_k-\lambda_j)N_{jk},
\end{equation}
where the elements of the matrices are represented in the basis $\{|j\rangle\}$.
This equality implies that we can always freely choose $N$ and determine $M$ accordingly in terms of $D$ and $N$.
When $\lambda_j+\lambda_k\neq0$, we have
\begin{equation}
    M_{jk} = \frac{2D_{jk}+i(\lambda_j-\lambda_k)N_{jk}}{\lambda_j+\lambda_k},
\end{equation}
which implies that $L_{jk} = (2D_{jk} + 2i\lambda_j N_{jk}) / (\lambda_j + \lambda_k)$. 
It then follows that
\begin{align}
    \tr(L^\dagger L \rho) 
    &= \sum_{j,k|\lambda_k>0} \lambda_k |L_{jk}|^2 \\
    &= \sum_{j,k|\lambda_k>0} \frac{4 \lambda_k}{(\lambda_j+\lambda_k)^2} |D_{jk} + i \lambda_j N_{jk}|^2 \\
    &= \sum_{j,k|\lambda_k>0} A_{jk} + \sum_{j,k|\lambda_j>0,\lambda_k>0} B_{jk},
\end{align}
where 
\begin{align}
    A_{jk} &:= \frac{4\lambda_k}{(\lambda_j+\lambda_k)^2}\left(|D_{jk}|^2 + \lambda_j^2 |N_{jk}|^2\right),\\
    B_{jk} &:= \frac{4i\lambda_j\lambda_k}{(\lambda_j+\lambda_k)^2} (D_{jk}^* N_{jk} - D_{jk} N_{jk}^*).
\end{align}
Since both $D$ and $N$ are Hermitian, we have $D_{jk}^*=D_{kj}$ and $N_{jk}^*=N_{kj}$, which implies that the matrix $B$ is antisymmetric as $B_{jk}=-B_{kj}$.
Therefore, 
\begin{equation}
    \tr(L^\dagger L \rho) = \sum_{j,k|\lambda_k>0} A_{jk} 
    \geq \sum_{jk|\lambda_j>0} \frac{4\lambda_k |D_{jk}|^2}{(\lambda_j+\lambda_k)^2}.
\end{equation}
The equality in the above inequality holds when all $N_{jk}$ vanishes, meaning that  $L$ is Hermitian. \qed

Now, let us consider the case where each $L_k(x)$ may be non-uniquely determined by $D_k(x)$ and $\rho(x)$ through Eq.~(\ref{eq:L_k}). 
Denote the Hermitian solution for $L_k(x)$ by $\tilde L_k(x)$ and define the matrix $\tilde G$ by $\tilde G_{kk'}=\Re\tr[\tilde L_k(x)^\dagger \tilde L_{k'}(x) \rho(x)]$.
Let $u$ be an arbitrary real vector. 
In terms of the above {\it Proposition} with $D=\sum_k u_k D_k(x)$ and $L=\sum_k u_k L_k(x)$, it can be shown that $u^\top G u \geq u^\top \tilde G u$, thus $G\geq\tilde G$. 
Suppose that $G$ is strictly positive, then $G\geq\tilde G$ implies $\tilde G^{-1} \geq G^{-1}$. 
Thus, the Hermitian $L_k(x)$ give the tightest lower bound in the Weiss-Weinstein family.

\section{Phase-estimation example} \label{app_c}

Here, we calculate the MMSE, the QWWB, the QCRB, and the QZZB for the first example in Sec.~\ref{sec:two_level}.
For a unitary sensing $U_x=\exp(-ixH)$ and a Gaussian prior distribution 
\begin{equation}
	p(x)=\frac{1}{\sqrt{2\pi}\sigma}\exp\left[-\frac{(x-\mu)^2}{2\sigma^2}\right],
\end{equation}
the MMSE of estimating $x$ is given by~\cite{Macieszczak2014}
\begin{equation}
    \Sigma_{\min} = \sigma^{2}-\sigma^{4}\mathcal{F}(\bar\rho,H),
\end{equation}
where $\mathcal{F}(\bar\rho,H)$ is the quantum Fisher information about a parameter $\theta$ in the parametric quantum state $U_\theta\bar\rho U_\theta^\dagger$, where $\bar\rho:=\int_{-\infty}^\infty\!dx\,p(x)U_x\rho U_x^\dagger$ with $\rho$ being the initial state. 
With the eigenvalue decomposition $\bar\rho=\sum_j \lambda_j |j\rangle\langle j|$, one has
\begin{equation}\label{supp:QFI}
    \mathcal{F}(\bar\rho,H) = \sum_{j,k|\lambda_j+\lambda_k>0}\frac{2(\lambda_j-\lambda_k)^2|H_{jk}|^2}{\lambda_j+\lambda_k}.
\end{equation}
In our example, the initial state is $|\psi\rangle=(|0\rangle+|1\rangle)/\sqrt2$ and the generator of the unitary sensing is $H=E|1\rangle\langle1|$, where $E$ is a positive number.
Then, the average state is given by
\begin{equation}
    \bar\rho=\frac12(|0\rangle\langle0|+|1\rangle\langle1|)+\frac\gamma2(|0\rangle\langle1|+|1\rangle\langle0|) 
\end{equation}
with $\gamma:=\exp(-E^2\sigma^2/2)$.  
The eigenvalues and eigenvectors of $\bar\rho$ are $(1\pm\gamma)/2$ and $(|0\rangle\pm|1\rangle)/\sqrt2$ respectively.
It then follows from Eq.~(\ref{supp:QFI}) that $\mathcal{F}(\bar\rho,H)=\gamma^2E^2$, which implies 
\begin{equation}
    \Sigma_{\min}=\sigma^2-\sigma^4 E^2 \exp(-E^2\sigma^2).
\end{equation}

To obtain the QWWB, one only needs $g_\mathrm{c}(s,h)=\exp[-h^2s(1-s)/(2\sigma^2)]$ and $z(h)= (1+e^{-iEh})/2$, with which the QWWB is give by
\begin{widetext}
\begin{equation}
    \Sigma_\mathrm{W}(s,h) = \frac{h^2 g_\mathrm{c}(s,h)^2 |z(h)|^4}
    {[g_\mathrm{c}(2s,h)+g_\mathrm{c}(2-2s,-h)]|z(h)|^2 - 2 g_\mathrm{c}(s,2h)\Re\,z(h)^2 z(2h)^*}. 
\end{equation}
\end{widetext} 
Taking $s=1/2$ for simplicity, we obtain the QWWB optimized over $h$ as follows: 
\begin{equation}
    \Sigma_\mathrm{W} = \sup_h \frac{h^2\exp(-\frac{h^2}{4\sigma^2})\cos(\frac{hE}{2})^2}{2-2\exp(-\frac{h^2}{2\sigma^2})\cos(hE)}.
\end{equation}
After some algebras, the QCRB is given by 
\begin{equation}
    \Sigma_\mathrm{C}=\frac{1}{1/\sigma^2+E^2},
\end{equation}
and the QZZB is given by 
\begin{equation}
    \Sigma_\mathrm{Z}=\frac12\int_0^{+\infty}\!dh\, h\,\mathrm{erfc}\left(\frac{h}{2\sqrt{2}\sigma}\right) [1-\sqrt{1-|z(h)|^2}],
\end{equation}
where $\mathrm{erfc}(x)=(2/\sqrt\pi)\int_x^{+\infty}\!dt\,e^{-t^2}$ is the complementary error function.

\bibliography{qwwb}

\end{document}